\documentclass{article}
\usepackage{spconf,amsmath,graphicx}

\usepackage{enumitem}
\setlist{nosep, leftmargin=14pt}

\usepackage{mwe} 

\usepackage{amsmath,amssymb,amsfonts}
\usepackage{algorithmic}
\usepackage{graphicx}
\usepackage{textcomp}
\usepackage{xcolor}
\usepackage{soul}

\title{A Generative Imputation Method for Multimodal Alzheimer\textquotesingle{}s Disease Diagnosis}

\name{Reihaneh Hassanzadeh$^{\star \dagger}$ \qquad Anees Abrol$^{\dagger}$ \qquad Hamid Reza Hassanzadeh$^{\ast}$ \qquad Vince D. Calhoun$^{\star \dagger}$}

\address{$^{\star}$ School of Electrical and Computer Engineering, Georgia Institute of Technology, Atlanta, GA, USA \\
    $^{\dagger}$ Tri-institutional Center for Translational Research in Neuroimaging and Data Science (TReNDS), \\
    Atlanta, GA, USA \\
    $^{\ast}$ Courtesy Faculty Appointment, College of Pharmacy, University of Florida, Gainesville, FL, USA \\
    }

\begin{document}
%
\maketitle
\begin{abstract}
Multimodal data analysis can lead to more accurate diagnoses of brain disorders due to the complementary information that each modality adds. However, a major challenge of using multimodal datasets in the neuroimaging field is incomplete data, where some of the modalities are missing for certain subjects. Hence, effective strategies are needed for completing the data. Traditional methods, such as subsampling or zero-filling, may reduce the accuracy of predictions or introduce unintended biases. In contrast, advanced methods such as generative models have emerged as promising solutions without these limitations. In this study, we proposed a generative adversarial network method designed to reconstruct missing modalities from existing ones while preserving the disease patterns. We used T1-weighted structural magnetic resonance imaging and functional network connectivity as two modalities. Our findings showed a 9\% improvement in the classification accuracy for Alzheimer’s disease versus cognitive normal groups when using our generative imputation method compared to the traditional approaches.
\end{abstract}
\section{Introduction}

Alzheimer's disease (AD) is a neurodegenerative brain disorder that affects an increasing number of individuals worldwide \cite{vos2016global}. AD is characterized by a progressive decline in cognitive and functional abilities, such as memory loss, confusion, and difficulty in problem-solving \cite{breijyeh2020comprehensive}. There is currently no cure for AD; however, early diagnosis and intervention can offer the best possible care for those affected. Hence, the development of tools that can provide accurate predictions is of great importance.

Neuroimaging tools, such as structural magnetic resonance imaging (sMRI) and functional MRI (fMRI), are promising approaches to predicting AD. sMRI detects changes in the structure (i.e., brain atrophy) and fMRI detects changes in the brain activation patterns (e.g., reduced connectivity between functional brain networks) that are associated with AD. While each of these modalities provides unique insight into AD, multimodal analysis may lead to more accurate diagnoses of brain disease benefiting from complementary information offered by each modality \cite{venugopalan2021multimodal}. However, a prevalent challenge of using multimodal data in the neuroimaging field is the missing modalities for certain subjects, demanding effective strategies for data completion.

Traditional methods, such as subsampling or zero-filling \cite{venugopalan2021multimodal}, may reduce the accuracy of predictions or introduce unintended biases. In contrast, advanced methods such as generative models have emerged as promising solutions to generate missing data without these limitations. Most existing studies used generative models to transfer one modality into another while the source and target modalities are similar in nature or have the same dimensionality \cite{zhou2019latent,gao2021task,tiago2023domain,dar2019image,yuan2020unified}. Yet, the more challenging task occurs when modalities are in different spaces and differ in the types of brain-related information they provide remains largely unaddressed. Furthermore, the majority of the previous works have focused on a controlled-case dataset \cite{ellis2022evaluation,cao2020auto,cheng2021research}, whereas applying generative methods to disorder-specific datasets, with their potential to identify disease-related patterns, is more complex and remains largely unexplored.
 
In this study, as a novel endeavor, we utilized generative learning to synthesize structural and functional brain imaging data from each other. More specifically, we employed a cycle-generative adversarial network (Cycle-GAN) \cite{zhu2017unpaired,hassanzadeh2024cross} to transform one-dimensional functional network connectivity (FNC) maps \cite{dennis2014functional}, derived from fMRI, into three-dimensional T1 images and vice versa in the context of Alzheimer's disease. We imputed the missing samples with their corresponding generated ones in the multi-modal classification of Alzheimer's disease (AD) versus cognitively normal (CN) subjects. Our generative imputation method resulted in a 9\% improvement in classification accuracy compared to multiple baselines established for comparative analysis.

\begin{figure*}[thpb]
    \begin{center}
        \includegraphics[width=1\textwidth]{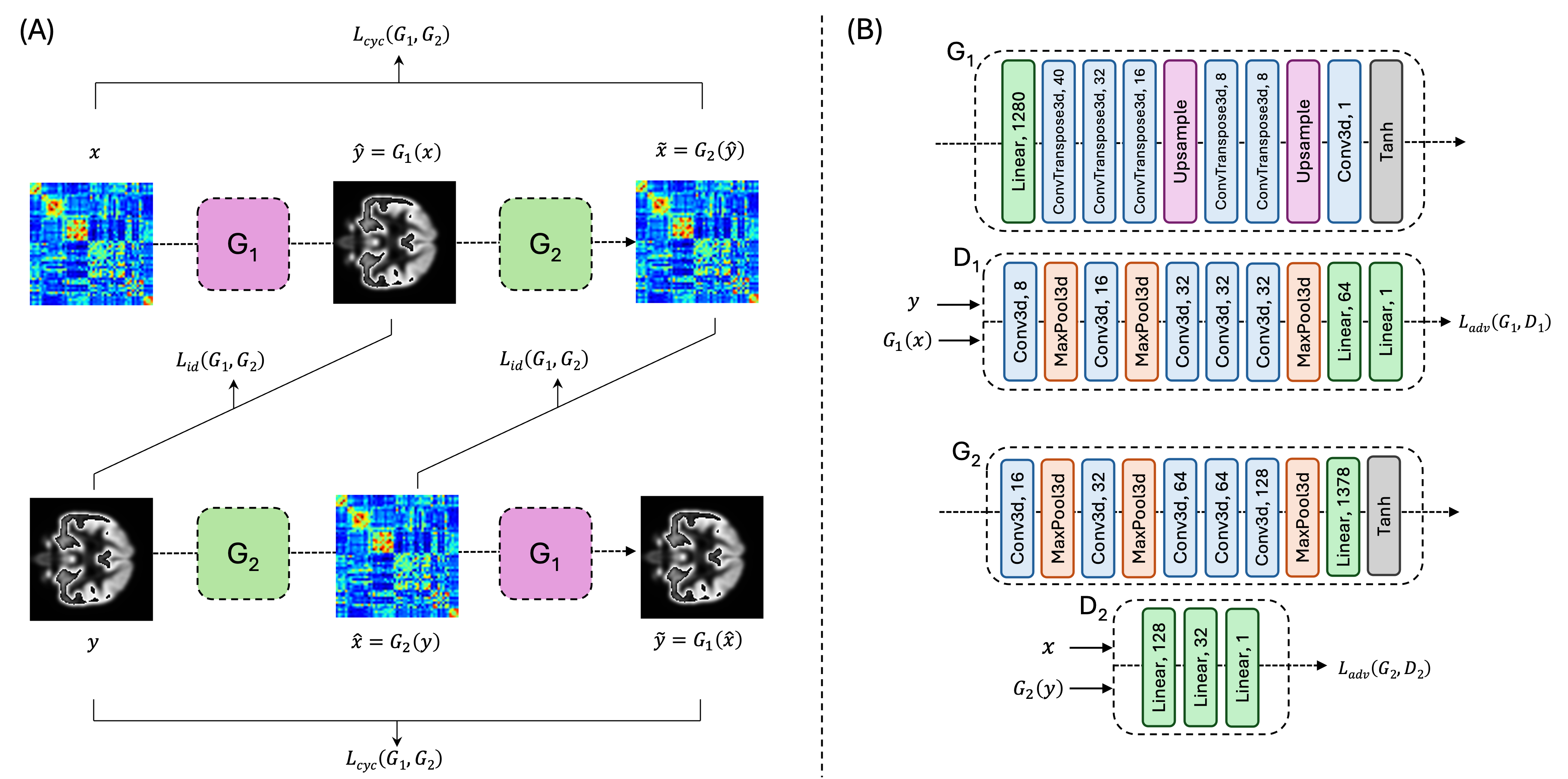}
        \centering
        \caption{Generative model architecture. The model includes two generators, \( G_1 \) and \( G_2 \), which transform FNC maps to T1 images and vice versa, and two discriminators, \( D_1 \) and \( D_2 \), which distinguish real samples from generated ones. Part (A) shows the data flow and loss functions, while part (B) details the architecture of each network component.
    }
        \label{fig:framework}
    \end{center}
\end{figure*}

\section{Methodology}
\subsection{Generative Model}

We employed a cycle-GAN network to translate the input data between the FNC and T1 domains. The original cycle-GAN learns the underlying distribution of the two domains and maps the distribution to each other using an unpaired data transition and under an unsupervised setting. In this study, we also incorporated weak supervision when paired data were available. Our proposed model, as illustrated in Fig. \ref{fig:framework}, consists of two generators, $G_1$ and $G_2$, to translate data between the two domains, and two discriminators, $D_1$ and $D_2$, to distinguish the real samples from the generated ones. 

The architecture of each network is depicted in Fig. \ref{fig:framework}.B. The generator $G_1$ transforms FNC maps, each with a size of 1378, into T1 images, each with dimensions of 121x145x121, via a sequence of five three-dimensional transposed convolution layers, each followed by batch normalization, and a final convolutional layer followed by a tanh activation layer. To match the final output size with the real T1 images, the outputs of the third and last layers were upsampled using an upsampling layer. The generator $G_2$ converts T1 images into FNC maps using five three-dimensional convolutional layers, each followed by batch normalization and a max-pooling layer. A linear layer and a tanh activation function were applied to the outputs of the final convolutional layer to produce FNC maps of the original size. The discriminator $D_1$ distinguishes between real and fake T1 images using five three-dimensional convolutional layers, some followed by a max-pooling layer, and the discriminator $D_2$ differentiates between real and fake FNC maps using three linear layers, as shown in the figure.

The adversarial loss functions used for the generators and discriminators are defined as follows:
\begin{equation}
    \mathcal{L}_{adv}(G_1, D_1) = 
    \begin{aligned}[t]
        &\mathbb{E}_{y \sim p_{\text{data}}(y)} \left[\left(D_1(y) - 1\right)^2\right] \\
        &+ \mathbb{E}_{x \sim p_{\text{data}}(x)} \left[\left(D_1(G_1(x))\right)^2\right],
    \end{aligned}
\end{equation}
\begin{equation}
    \mathcal{L}_{adv}(G_2, D_2) = 
    \begin{aligned}[t]
        &\mathbb{E}_{x \sim p_{\text{data}}(x)} \left[\left(D_2(x) - 1\right)^2\right] \\
        &+ \mathbb{E}_{y \sim p_{\text{data}}(y)} \left[\left(D_2(G_2(y))\right)^2\right].
    \end{aligned}
\end{equation}

To ensure the generators accurately map between modalities, we used the cycle consistency loss:
\begin{equation}
    \mathcal{L}_{cyc}(G_1, G_2) = 
    \begin{aligned}[t]
        &\mathbb{E}_{x \sim p_{\text{data}}(x)} \left[\left\| G_2(G_1(x)) - x \right\|_1\right] \\
        &+ \mathbb{E}_{y \sim p_{\text{data}}(y)} \left[\left\| G_1(G_2(y)) - y \right\|_1\right].
    \end{aligned}
\end{equation}

Additionally, we incorporated weak supervision by using identity loss for paired data:
\begin{equation}
    \mathcal{L}_{id}(G_1, G_2) = 
    \begin{aligned}[t]
        &\sum_{(x,y) \in \mathcal{P}} \left[\left\| G_1(x) - y \right\|_1\right] \\
        &+\sum_{(x,y) \in \mathcal{P}} \left[\left\| G_2(y) - x \right\|_1\right],
    \end{aligned}
\end{equation}
where $\mathcal{P}$ denotes the set of paired data.

The overall objective function for the model combines these losses with weight parameters $\lambda_1$ and $\lambda_2$ became:
\begin{equation}
    \mathcal{L}(G_1, G_2, D_1, D_2) = 
    \begin{aligned}[t]
        &\mathcal{L}_{adv}(G_1, D_1) \\
        &+ \mathcal{L}_{adv}(G_2, D_2) \\
        &+ \lambda_1 \mathcal{L}_{cyc}(G_1, G_2) \\
        &+ \lambda_2 \mathcal{L}_{id}(G_1, G_2),
    \end{aligned}
\end{equation}

\subsection{Multi-Modal Classifier}
\begin{figure}[thpb]
    \includegraphics[width=1\columnwidth]{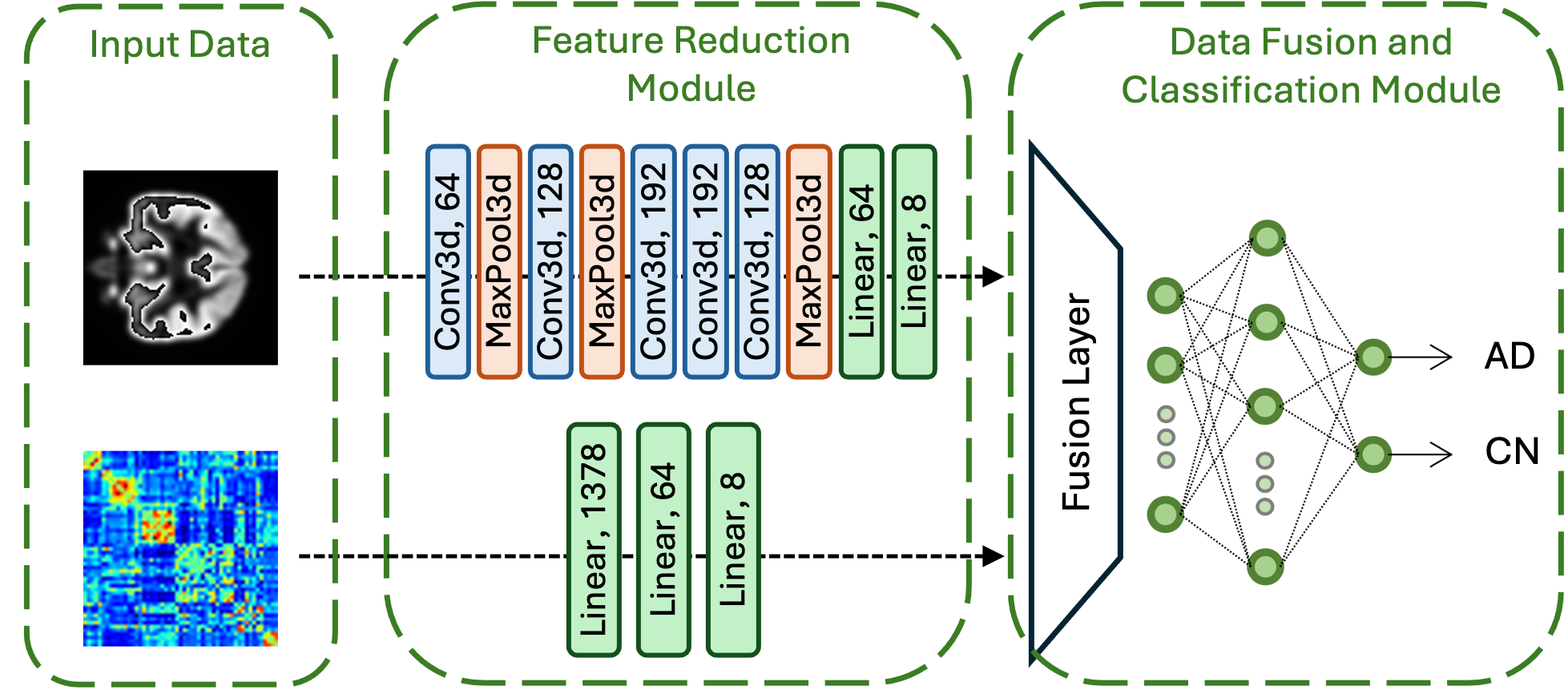}
    \centering
    \caption{Multi-modal classification network. The network takes FNC maps and T1 images as inputs and processes them through several convolutional and fully connected layers to extract high-level features. The extracted features were then fused together to form a comprehensive feature layer, which was used for the final classification task.
}
    \label{fig:classification}
\end{figure}

As illustrated in Fig. \ref{fig:classification}, the multi-modal classification network takes FNC maps and T1 images as inputs and processes them through several convolutional and fully connected layers to extract high-level features. For FNC maps, we used three fully connected layers with sizes 1378, 64, and 8 to extract high-level features. For T1 images, we used a sequence of five 3D-CNN layers with channel sizes 64, 128, 192, 192, and 128 followed by two fully connected layers of sizes 64 and 8 for feature extraction. The features extracted were then fused together (i.e., concatenated) to form a comprehensive feature layer with a size of 16. Finally, we added several fully connected layers on top of the feature layer for the final classification task.

\subsection{Implementation Details}
We employed a five-fold cross-validation method stratified on diagnosis to split the data into training and test sets. We trained and evaluated our generative and classification models using the training set, and tested the classification model on the test set. The test set was kept unseen during the learning process and was only used to report the final classification accuracy. The generative and classification models were trained for 300 and 200 epochs, respectively. In the generative model, we used a batch size of 32 and utilized an Adam optimizer starting with a learning rate of 0.05, which decreased by 0.9. We selected $\lambda_1=10$ and $\lambda_2=40$ via a grid search within the ranges of 1, 5, 10, 20, and 40. To further reduce the possibility of mode collapse \cite{goodfellow2020generative}, we updated the discriminators using the latest 50 generated samples which were kept in a buffer. In the classification model, we used a batch size of 16 and an Adam optimizer, with a starting learning rate selected through a grid search among 0.01, 0.001, 0.0001, and 0.00001, which decayed by 0.98 each epoch.

\subsection{Dataset}
In this study, we used 2923 brain imaging samples of 986 subjects from the Alzheimer's Disease Neuroimaging Initiative (ADNI)\cite{jack2008alzheimer}. Table \ref{table:data-dist} displays the distribution of data across diagnosis and modality groups. 
Although 2910 T1 images were included in the dataset, 414 FNC maps were available. 

An independent component analysis (ICA) pipeline \cite{du2020neuromark}, NeuroMark, was used to decompose the fMRI data into 53 components. These components were functionally grouped into seven domains: auditory (AU), subcortical (SC), sensorimotor (SM), visual (VI),  default mode (DM), cognitive control (CC), and cerebellar (CB). To generate FNC features, the Pearson correlation between each two of the 53 components was calculated, resulting in 1374 FNC features. 

\begin{table}[h!]
\caption{Data Distribution.}
\centering
\begin{tabular}{|c|c|c|c|}
\hline
Diagnosis & \multicolumn{3}{c|}{Sample Size} \\ \cline{2-4} 
                           & FNC   & T1    & FNC and T1      \\ \hline
CN                         & 207   & 1446   & 207            \\ \hline
AD                         & 207   & 1465   & 195            \\ \hline
\end{tabular}
\label{table:data-dist}
\end{table}

\section{Experimental Results}

\subsection{Generative Model Performance}
\begin{figure}[htbp]
    \begin{center}
        \includegraphics[width=1\columnwidth]{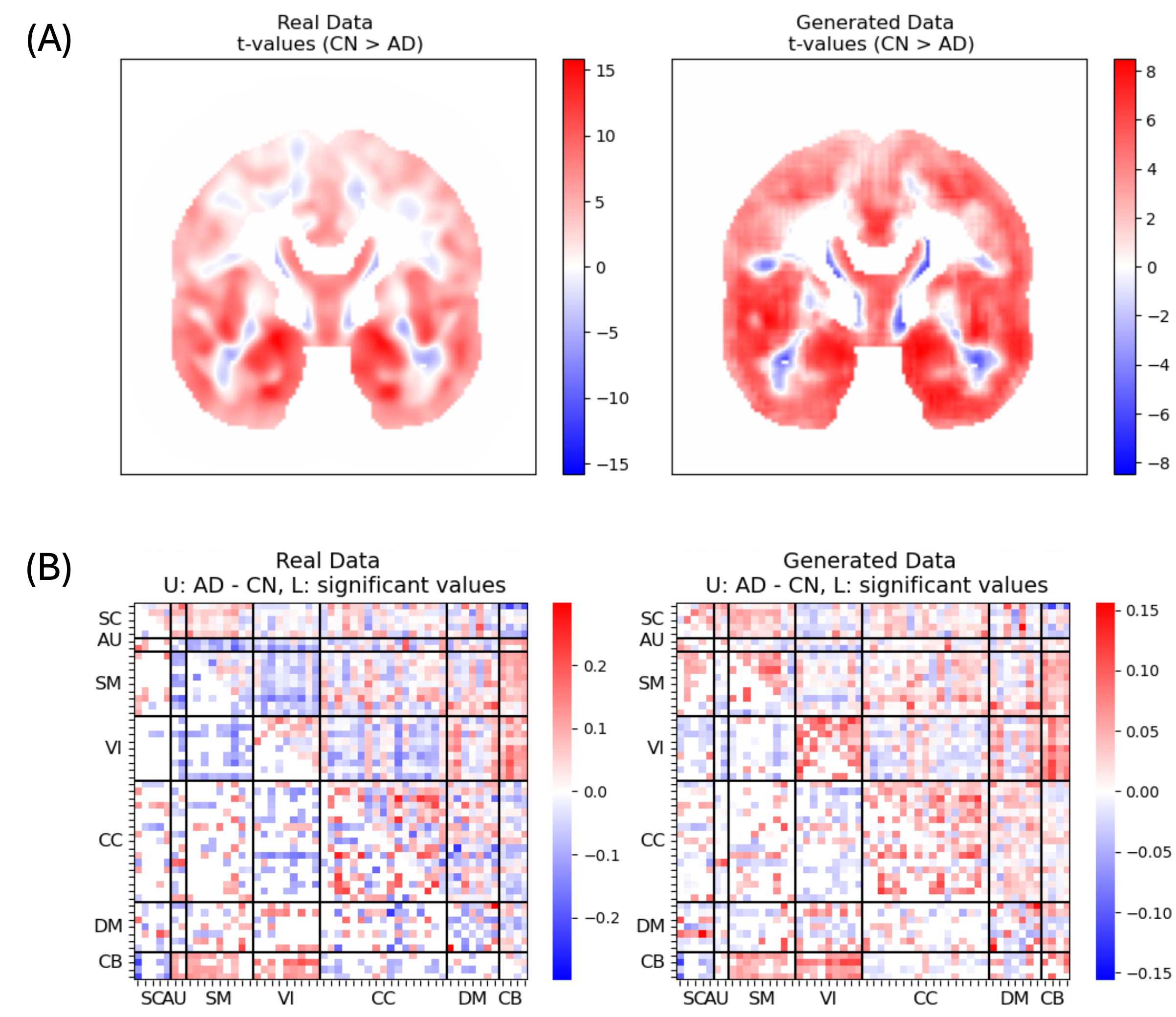}
        \caption{(A) t-values between T1 images of AD and CN. (B) Group differences between FNC maps of AD and CN. U and L indicate the Upper and Lower triangular matrix, respectively.}
        \label{fig:group_diff}
    \end{center}
\end{figure}

We adopted the structural similarity index measure (SSIM) and the Peak Signal-to-Noise Ratio (PSNR) between the real T1 images and their corresponding generated T1 images, as well as the Mean Squared Error (MSE) and Pearson correlation between the real FNC features and the generated ones, to evaluate the performance of our generative model. Our results showed an SSIM of 0.89$\pm$0.003, a PSNR of 24.915$\pm$0.372, an MSE of 0.083 $\pm$0.002, and a Pearson correlation of 0.71$\pm$0.004. Furthermore, Fig. \ref{fig:group_diff} visually shows that the generated data could capture the diagnostic patterns in the real data. More specifically, Fig. \ref{fig:group_diff}.A shows the t-values between T1 images of AD and CN, suggesting a similar atrophy in the generated images compared to the real ones of Alzheimer’s patients, in particular the hippocampal and other temporal regions. Fig. \ref{fig:group_diff}.B shows group differences in FNC maps of diagnosis groups and presents similar changes in functional connectivity, for example, increases between CB and SM \cite{mckhann2011diagnosis} and between CB and VI networks and decreases between CB and SC networks \cite{hassanzadeh2022classification}, and between AU and VI networks.

\subsection{Multi-Modal Classification Performance}

\begin{figure}[htbp]
    \begin{center}
        \includegraphics[width=1\columnwidth]{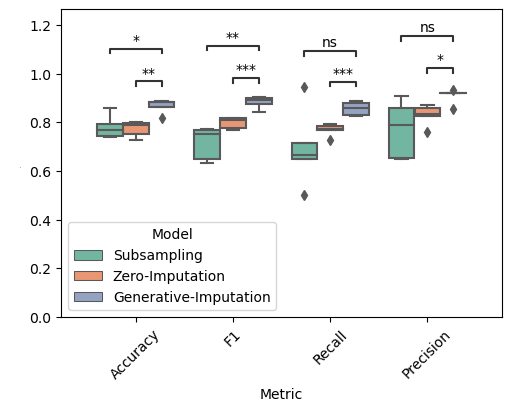}
        \caption{Classification Performance of AD vs. CN. Statistical significance was determined using t-tests, with the following p-value annotations: $\text{ns}: p > 0.05$, $\text{*}: 0.01 < p \leq 0.05$, $\text{**}: 0.001 < p \leq 0.01$, and $\text{***}: 0.0001 < p \leq 0.001$.}
        \label{fig:performance}
    \end{center}
\end{figure}

Using the real T1 and FNC data along with the generated data we trained a multi-modal classification of AD vs. CN and measured the performance of the model with the accuracy, precision, recall, and F1 score. Moreover, we compared the performance of the model with the following baselines: 1. subsampling, where the input data includes only the data for which both modalities are available; and 2. zero-imputation, where the missing modality is replaced with zeros. Fig. \ref{fig:performance} shows a summary of the performance of each approach. According to the results, our generative-imputation approach achieved an accuracy of 86.87\%$\pm$2.9 and outperformed the subsampling and zero-imputation approaches by 8.6\% and 9.4\%, respectively. Furthermore, our proposed approach attained an F1 score of 0.88, a recall of 0.86, and a precision of 0.91, all of which were superior to the baselines.

\section{Conclusion}
In this study, we explored the capability of generative models for the brain function-structure translation within the context of Alzheimer's disease. We developed a Cycle-GAN adapted to our data to synthesize functional connectivity maps and T1 images from each other. Our findings suggested that this approach could learn distinctive brain patterns associated with Alzheimer's disease. We then applied our generative method to address the missing modality data by integrating the generated samples into a multi-modal classification model. This generative imputation method resulted in a 9\% improvement in classification accuracy compared to the baselines. As an interesting future direction, one could explore the application of other generative models, such as diffusion models, in multi-modal disease diagnosis.



\bibliographystyle{IEEEbib}
\bibliography{bibliography}

\end{document}